\documentclass[11pt]{article}
\usepackage{epsfig}
\usepackage{graphicx}
\usepackage{amsmath}
\usepackage{hhline}
\usepackage{amssymb}
\usepackage{times}
\usepackage{cite}
\usepackage{array}
\usepackage{multirow}

\usepackage{color}
\definecolor{rltred}{rgb}{0.75,0,0}
\definecolor{rltgreen}{rgb}{0,0.5,0}
\definecolor{rltblue}{rgb}{0,0,0.75}

\newlength{\dinwidth}
\newlength{\dinmargin}
\begin{document}
\begin{titlepage}

\noindent
Date:        \today       \\
                
\vspace{2cm}

\begin{center}
\begin{Large}

{\bf The evolution of $\sigma^{\gamma P}$ with coherence length}

\vspace{2cm}

Allen Caldwell \\
Max Planck Institute for Physics (Werner-Heisenberg-institut) \\
Munich, Germany

\end{Large}
\end{center}

\vspace{2cm}

\begin{abstract}

 Assuming the form $\sigma^{\gamma P} \propto l^{\lambda_{\rm eff}}$ at fixed $Q^2$ for the behavior of the virtual-photon proton scattering cross section, where $l$ is the coherence length of the photon fluctuations, it is seen that the extrapolated values of $\sigma^{\gamma P}$ for different $Q^2$ cross for $l\approx 10^8$~fm. It is argued that this behavior is not physical, and that the behavior of the cross sections must change before this coherence length $l$ is reached.  This could set the scale for the onset of saturation of parton densities.

\end{abstract}
\end{titlepage}

\section{Introduction}

According to quantum field theory, the microscopic world is a dynamic environment where short-lived states are constantly being created and annihilated.  Increasing the resolution of a probe used to sense this environment, we see ever more structure until, we anticipate, at some small enough distance scale we see universal distributions.  
It has already been observed that the energy dependence of the scattering cross section of high energy hadrons becomes universal~\cite{ref:DL} and that the size of the cross section only depends on the number of valence quarks in the scattering particles and the center-of-mass energy.  We  study in this paper the energy dependence of scattering cross sections for virtual photons on protons, where the interaction cross section is dominated by the strong interaction.  We anticipate that in a high energy limit, the scattering cross sections of virtual photons will also achieve a universal behavior since the interaction will have as source the photon fluctuating into a quark-antiquark pair.

 In deep inelastic scattering of electrons on protons at HERA, the strong increase in the proton structure function $F_2$ with decreasing $x$ for fixed, large, $Q^2$ is interpreted as an increasing density of partons in the proton, providing more scattering targets for the electron.  This interpretation relies on choosing a particular reference frame to view the scattering - the Bjorken frame.  In the frame where the proton is at rest, it is the state of the photon or weak boson that differs with varying kinematic parameters.  For the bulk of the electron-proton interactions, the scattering process involves a photon, and we can speak of different states of the photon scattering on a fixed proton target.  What is seen is that the photon-proton cross section rises quickly with photon energy $\nu$ for fixed virtuality $Q^2$~\cite{ref:sigmagp}.  We interpret this as follows: as the energy of the photon increases, time dilation allows shorter lived fluctuations of the photon to become active in the scattering process, thereby increasing the scattering cross section.  We use the concept of coherence length of the short-lived photon states to analyze the behavior of the photon-proton cross section.
 
 This analysis is an update of the work reported in ~\cite{ref:caldwell} and has been inspired by the space-time picture of Gribov~\cite{ref:Gribov}.

\section{Coherence Length}

The Hand convention~\cite{ref:Hand} is used to define the photon flux 
yielding the relation:
\begin{equation}
\label{eq:F2}
F_2^P(x,Q^2)=\frac{Q^4(1-x)}{4\pi^2\alpha (Q^2+(2xM_P)^2)}\sigma^{\gamma P} 
\end{equation}
where $F_2^P$ is the proton structure function, $\alpha$ is the fine structure constant and $M_P$ is the proton mass. $Q^2$ and $x$ are the standard kinematic variables used to describe deep inelastic scattering.

\begin{figure}[hbpt]
\begin{center}
   \includegraphics[width=0.4\textwidth]{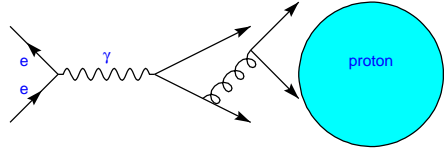}
\caption{\it Photon fluctuations scattering on the proton in the proton rest frame.}
\label{fig:proton}
\end{center}
\end{figure}

The behavior of $\sigma^{\gamma P}$ is studied in the proton rest frame in terms of the coherence length~\cite{ref:LES} of the photon fluctuations, $l$, and the virtuality, $Q^2$.  The physical picture is given in Fig.~\ref{fig:proton}, where the electron acts as a source of photons, which in turn acts as a source of quarks, antiquarks and gluons.  
We expect the scattering cross section to increase with coherence length since the photon has more time to develop structure. In the proton rest frame, the proton is a common scattering target for the incoming partons independent from the values of $Q^2$ and $l$.

We recall the definition of the coherence length, $l$~\cite{ref:BK}:
$$l \equiv \frac{\hbar c}{\Delta E}$$
where $\Delta E$, the change in energy of the photon as it fluctuates into a system of quarks and gluons, is given by
\begin{eqnarray}
\Delta E & \approx & \frac{m^2+Q^2}{2\nu}
\end{eqnarray}
where $\nu$ is the photon energy and $m$ is the mass of the partonic state.  For $Q^2 \gg m^2$, we have
\begin{equation}
\label{eq:cohere}
\Delta E \approx \frac{Q^2}{2\nu} 
\end{equation}
 and 
 \begin{equation}
 \label{eq:coherence}
l \approx  \frac{2 \nu \hbar c}{Q^2} \approx \frac{\hbar c}{x M_P}\;\; .
\end{equation}
Note that $M_P$ only appears because of the definition of $x$ and the value of the coherence length is not dependent on the proton mass.  The coherence length increases linearly with the photon energy and inversely proportional to the virtuality. In the limit $Q^2 \rightarrow 0$, the dependence of the coherence length will be modified since the $m^2$ term cannot be ignored.  We will therefore limit our study of the limiting behavior of the cross section to the regime $Q^2\geq 1$~GeV$^2$ where we expect that Eq.~\ref{eq:cohere} should be valid.

\section{Data Sets}
\label{sec:data}

The analysis relies primarily on the recently released combination of HERA electron-proton neutral-current scattering data from the ZEUS and H1 experiments~\cite{ref:HERAcombined}.  This is complemented by the small-$x$ data from the E665~\cite{ref:E665} and NMC~\cite{ref:NMC} experiments.  We use data in the kinematic range: $0.01<Q^2<500$~GeV$^2$, $x<0.01$ and $0.01<y<0.8$ for a global analysis and a more restricted range when studying the extrapolated behavior of the cross sections.  The HERA data are presented in the form of reduced cross sections
$$\sigma_{ \rm red}  = \frac{xQ^4}{2 \pi \alpha^2 Y_{+}} \frac{d \sigma^2}{dx dQ^2} \approx \left[F_2 - \frac{y^2}{Y_+}F_L \right] $$
where $Y_+= 1+(1-y)^2$.
The structure function $F_3$ can be ignored in the $(x,Q^2)$ range selected for the analysis. We calculate $F_2$ from the reduced cross section assuming $R=F_L/(F_2-F_L)=0.25$, and limit the values of $y$ to be smaller than $0.8$ to control the possible error due to this assumption.  The values of $x$ are restricted to be less than $0.01$ to ensure that we are dealing with large coherence lengths compared to the proton size.  The total experimental uncertainties are used (statistical and systematic added in quadrature on a point-by-point basis).  The use of correlated systematics was investigated but found not to give significant differences for this analysis.  In any case, we are looking more for qualitative behavior.

In total, 45 E665 data points, 13 NMC data points, 23 HERA $e^-P$ data points, and (115, 160, 75, 276)  HERA $e^+P$ data points with $E_p=(460,575,820,920)$~GeV satisfied the selection criteria.

\subsection{Bin-centering and further analysis}
\label{sec:center}
The data from the different experiments are reported in some cases at fixed values of $x$ and varying $Q^2$, in other cases at fixed $Q^2$ and varying $x$, or at fixed $Q^2$ and varying $y$.  This is no problem for the model fitting procedure, but it does make data presentation cumbersome.  The data were therefore `bin-centered' by moving data to fixed values of $Q^2$ using a parametrization (see below) as follows:
\begin{equation}
\label{eq:bincenter}
F_2(Q^2_c,x) = \frac{F_2^{\rm pred}(Q^2_c,x)}{F_2^{\rm pred}(Q^2,x)}\frac{F_2(Q^2,x)}{S_{\rm expt}} 
\end{equation}
where  $S_{\rm expt} $ is a normalization factor for the experiment in question that resulted from the global fit.  The following $Q^2_c$ values have been used: $0.25, 0.4, 0.65, 1.2, 2, 3.5, 6.5, 10, 15, 22, 35, 45, 90, 120$~GeV$^2$.

\subsection{Functional form for global fit}
A global fit of all the data satisfying the selection criteria was performed to have a functional form for the bin-centering.  The functional form for the virtual-photon proton cross section was constructed using the following reasoning.

 \subsubsection{$Q^2$ dependence}
 For compact photon fluctuations, the maximum value of $\sigma^{\gamma P}$ is given by the size of the proton multiplied by $\alpha$, giving roughly $200$~$\mu$barn.  However, pQCD calculations have pointed to the property of `color transparency' for small dipoles~\cite{ref:transparency}, indicating that at large $Q^2$ the cross section should behave as $\sigma^{\gamma P} \propto 1/Q^2$.  I.e., the proton appears almost transparent for small dipoles with the cross section proportional to the size of the photon.  The photon state will have a maximum size which is expected to be set by the mass of the lightest vector meson.  An effective mass was used as a free parameter in the fits.  We therefore have the following factor in our parametrization:
  $$\sigma_0 \frac{M^2}{Q^2+M^2}$$
where $\sigma_0$ is expected to be a typical hadronic cross section (multiplied by $\alpha$).  This term has two free parameters $(M^2, \sigma_0)$.
 
 \subsubsection{$l$ dependence} 
As discussed in the introduction, we expect (and already know from  looking at previous data) that the cross sections will grow with increasing coherence length.  We also know that at very small $Q^2$, the energy dependence of the photon-proton cross section is close to that for hadron-hadron scattering~\cite{ref:sigmagp}, while the cross section rises more steeply with energy at larger $Q^2$.  The transition in the behavior sets in around $Q^2=1$~GeV$^2$~\cite{ref:sigmagp,ref:caldwell}.  We therefore use the following factor to model the $l$ dependence:
$$ \left( \frac{l}{l_0} \right)^{\epsilon_{\rm eff}}$$
with
\begin{eqnarray*}
\epsilon_{\rm eff} &=& \epsilon_0 \hskip 5.cm  Q^2 \leq Q^2_0 \\
\epsilon_{\rm eff} &=& \epsilon_1+\epsilon' \ln{(Q^2/Q^2_1)} \hskip 2.4cm  Q^2 \geq Q^2_1\\
\epsilon_{\rm eff} &=& \epsilon_0+ (\epsilon_1-\epsilon_0)\frac{ \ln(Q^2/Q^2_0)}{ \ln(Q^2_1/Q^2_0)}\hskip 1.3cm Q^2_0 < Q^2 < Q^2_1
\end{eqnarray*}

This term has five free parameters $(\epsilon_0, \epsilon_1, \epsilon', Q^2_0, Q^2_1)$.  The coherence length is measured in fm.

\subsubsection{Prediction for $F_2$}
\begin{figure}[hbpt]
\begin{center}
   \includegraphics[width=0.6\textwidth]{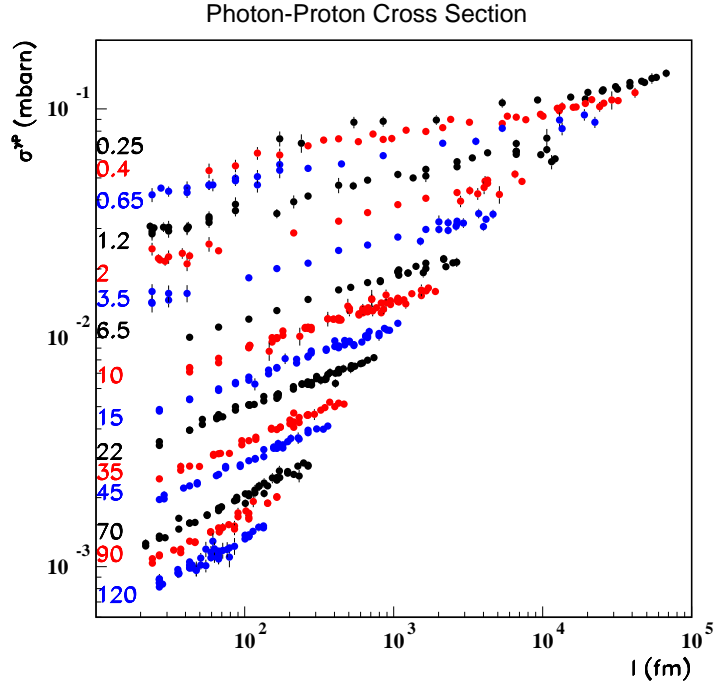}
\caption{\it Bin-centered photon-proton cross section data, shown in alternating colors for clarity. The values of $Q^2$ in GeV$^2$ for a given set of points are given at the left edge of the plot.}
\label{fig:bincentered}
\end{center}
\end{figure}

The photon-proton cross section is therefore fully parametrized with the eight parameters as
\begin{equation}
\label{eq:global}
\sigma^{\gamma p}= \sigma_0 \frac{M^2}{Q^2+M^2} \left( \frac{l}{l_0} \right)^{\epsilon_{\rm eff}(\epsilon_0, \epsilon_1, \epsilon', Q^2_0, Q^2_1)} \;  ,
\end{equation}
The prediction for $F_2$ (and subsequently the reduced cross section) is then achieved using Eq.~(\ref{eq:F2}), with the addition of one parameter for each data set to account for relative normalization uncertainties.  The bin-centered data are shown in Fig.~\ref{fig:bincentered}.  As is seen in the figure, the data cover several orders of magnitude in coherence length at the smaller values of $Q^2$, reducing to one order of magnitude at $Q^2=120$~GeV$^2$.  It is also clear that the slope of the cross section increases with increasing $Q^2$.

\section{Global Fit Results}
The Bayesian Analysis Toolkit (BAT)~\cite{ref:BAT} was used to extract the 15-dimensional posterior probability distribution of the parameters (the 8 parameters for the photon-proton cross section and additionally 7 normalizations) assuming Gaussian probability distributions for all data points.  The  prior probability distributions chosen are given in Table~\ref{tab:fit}.  For the parameters of the photon-proton cross section, the priors reflect the knowledge that exists on the parameters.  For the normalization parameters, the priors reflect the experimentally quoted uncertainties.

\begin{table}[htp]
\caption{Prior probability definitions for the parameters in the fit function and results of the global fit to the data. The notation $ x \sim \mathcal{G}(\circ|\mu,\sigma)$ means that $x$ is assumed to follow a Gaussian probability distribution centered on $\mu$ with variance $\sigma^2$. The parameter boundaries are also given. The parameter values at the global mode as well as the 68~\% smallest marginalized intervals from the global fit are given in the last two columns.}
\begin{center}
\begin{tabular}{|c|c|cc|}
\hline
Parameter & prior function & global mode &  68~\% interval  \\
\hline
$\sigma_0$ & $\sigma_0 \sim \mathcal{G}(\circ|0.07,0.02)  \;\; 0.01<\sigma_0<0.2$ & $0.062$ & $0.059-0.067$ \\
$M^2$ & $M^2\sim \mathcal{G}(\circ|0.75,0.5)  \;\; 0.1<M^2_0<2.0$ & $0.63$& $0.59-0.67$ \\
$l_0$ & $l_0 \sim \mathcal{G}(\circ|1.0,0.5)  \;\; 0.1<l_0<2.0$ & $1.6$ & $1.54-1.77$ \\
$\epsilon_0$ & $\epsilon_0 \sim \mathcal{G}(\circ|0.09,0.01)  \;\; 0.05<\epsilon_0<0.2$ & $0.106$ & $0.102-0.110$ \\
$\epsilon_1$ & $\epsilon_1 \sim \mathcal{G}(\circ|0.2,0.2)  \;\; 0.1<\epsilon_1<0.3$ & $0.156$ & $0.152-0.160$ \\
$\epsilon'$ & $\epsilon' \sim \mathcal{G}(\circ|0.05,0.02)  \;\; 0.0<\epsilon'<0.1$ & $0.052$ & $0.051-0.054$ \\
$Q^2_0$ & $Q^2_0 \sim \mathcal{G}(\circ|1.0,1.0)  \;\; 0.01<Q^2_0<2.0$ & $0.37$ & $0.33-0.41$ \\
$Q^2_1$ & $Q^2_1 \sim \mathcal{G}(\circ|3.0,3.0)  \;\; 2.0<Q^2_1<10.0$& $3.13$ & $2.96-3.36$  \\
$S_{\rm E665}$ & $S \sim \mathcal{G}(\circ|1.0,0.018)  \;\; 0.9<S_{\rm E665}<1.1$& $0.97$ & $0.958-0.982$  \\
$S_{\rm NMC}$ & $S \sim \mathcal{G}(\circ|1.0,0.025)  \;\; 0.9<S_{\rm NMC}<1.1$& $0.94$ & $0.093-0.096$ \\
$S_{\rm e^-p}$ & $S \sim \mathcal{G}(\circ|1.0,0.015)  \;\; 0.9<S_{\rm e^-p}<1.1$& $0.997$ & $0.988-1.006$ \\
$S_{\rm e^+p 460}$ & $S \sim \mathcal{G}(\circ|1.0,0.015)  \;\; 0.9<S_{\rm e^+p 460}<1.1$& $1.020$& $1.014-1.030$ \\
$S_{\rm e^+p 575}$ & $S \sim \mathcal{G}(\circ|1.0,0.015)  \;\; 0.9<S_{\rm e^+p 575}<1.1$ & $1.014$ & $1.008-1.024$\\
$S_{\rm e^+p 820}$ & $S \sim \mathcal{G}(\circ|1.0,0.015)  \;\; 0.9<S_{\rm e^+p 820}<1.1$ & $1.009$& $1.002-1.020$\\
$S_{\rm e^+p 920}$ & $S \sim \mathcal{G}(\circ|1.0,0.015)  \;\; 0.9<S_{\rm e^+p 920}<1.1$& $0.998$ & $0.992-1.008$ \\
\hline
\end{tabular}
\end{center}
\label{tab:fit}
\end{table}%

  The values of the parameters at the global mode are given in Table~\ref{tab:fit} together with the 68~\% smallest intervals for the 1D marginalized distributions.   The fit was satisfactory and was further used to bin-center the data points.  The posterior probability at the global mode corresponded to a $\chi^2$ value of 761 for 707 fitted data points.

\begin{figure}[hbpt]
\begin{center}
   \includegraphics[width=0.45\textwidth]{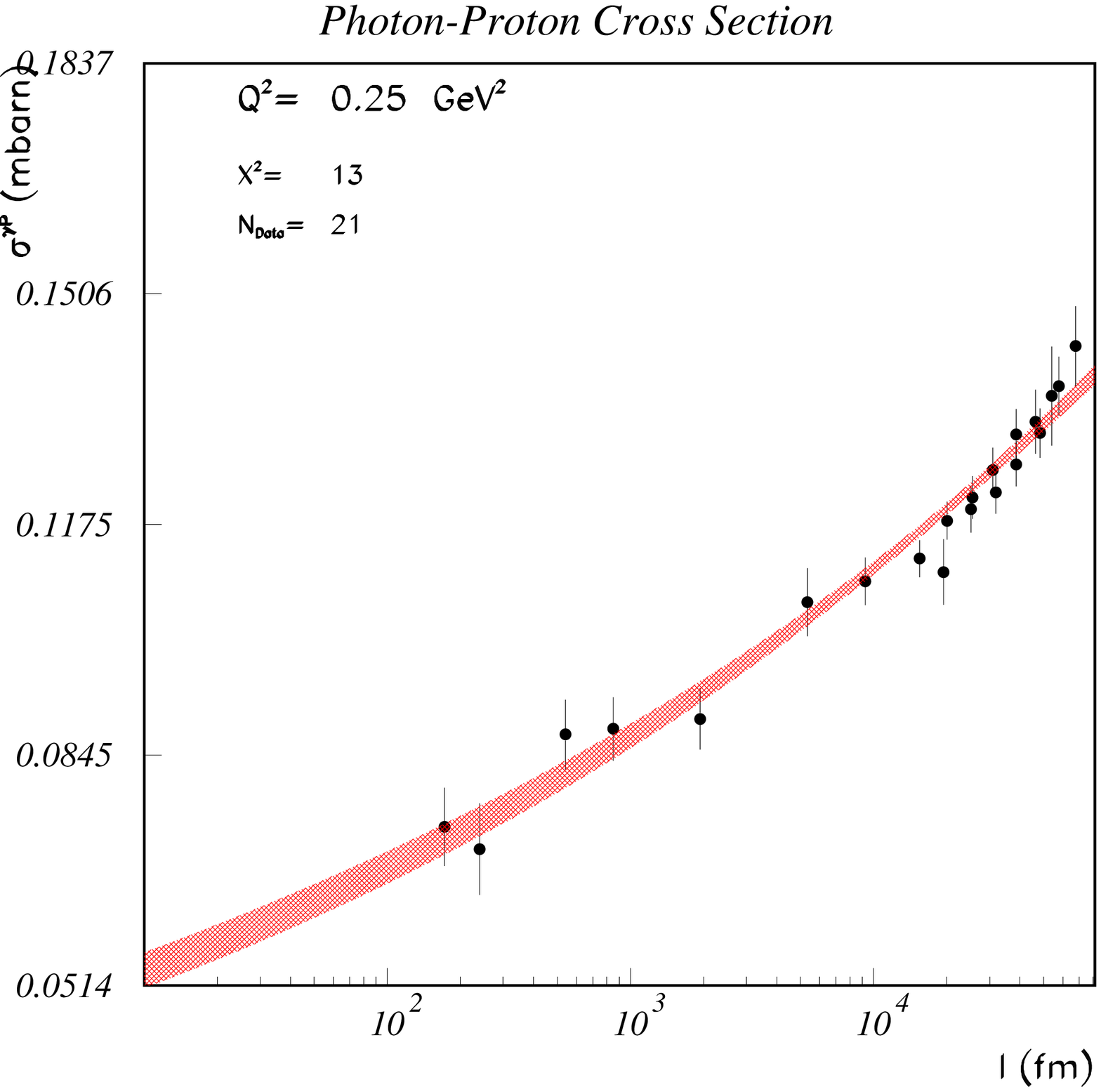}
   \includegraphics[width=0.45\textwidth]{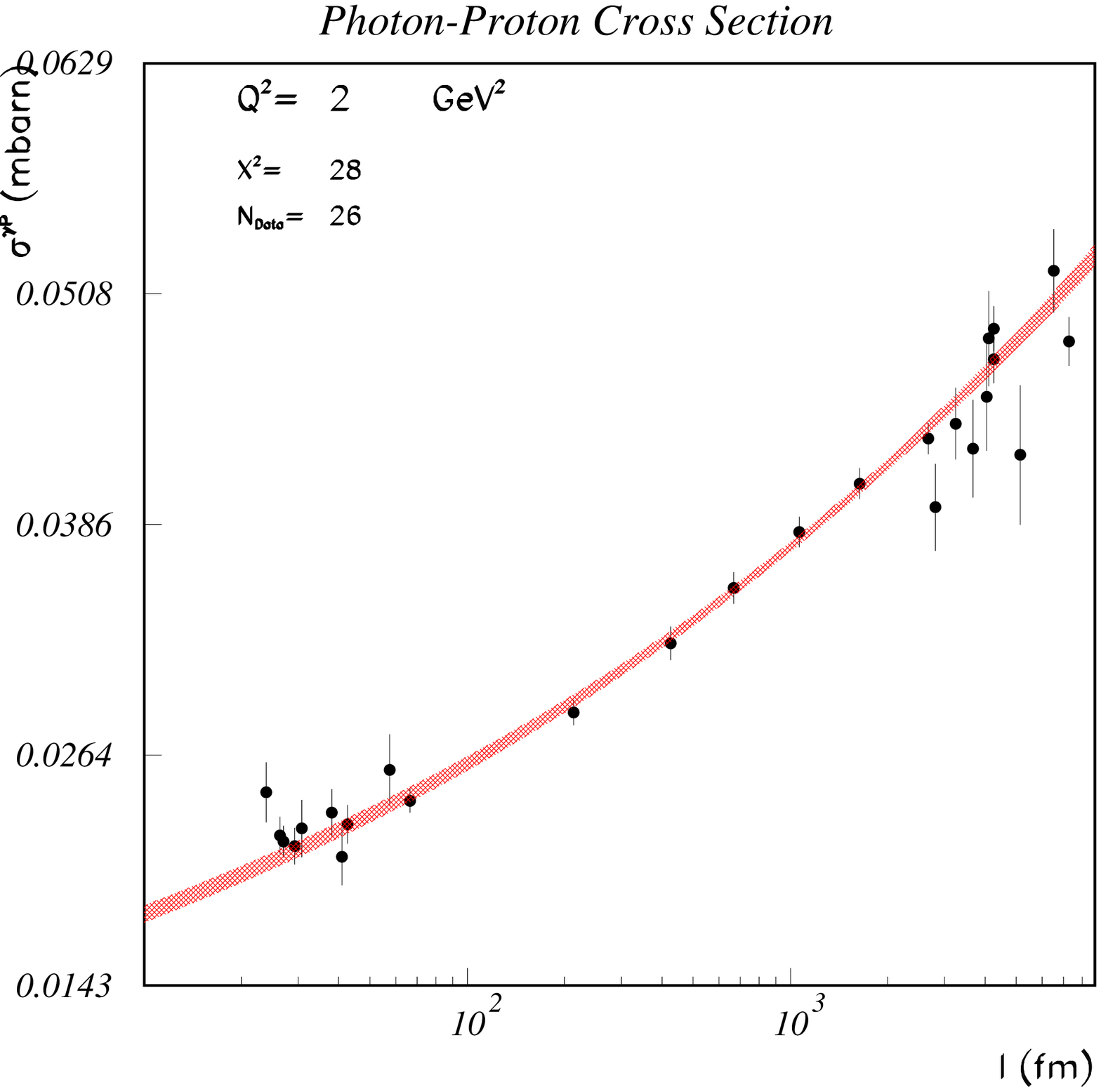}
   \includegraphics[width=0.45\textwidth]{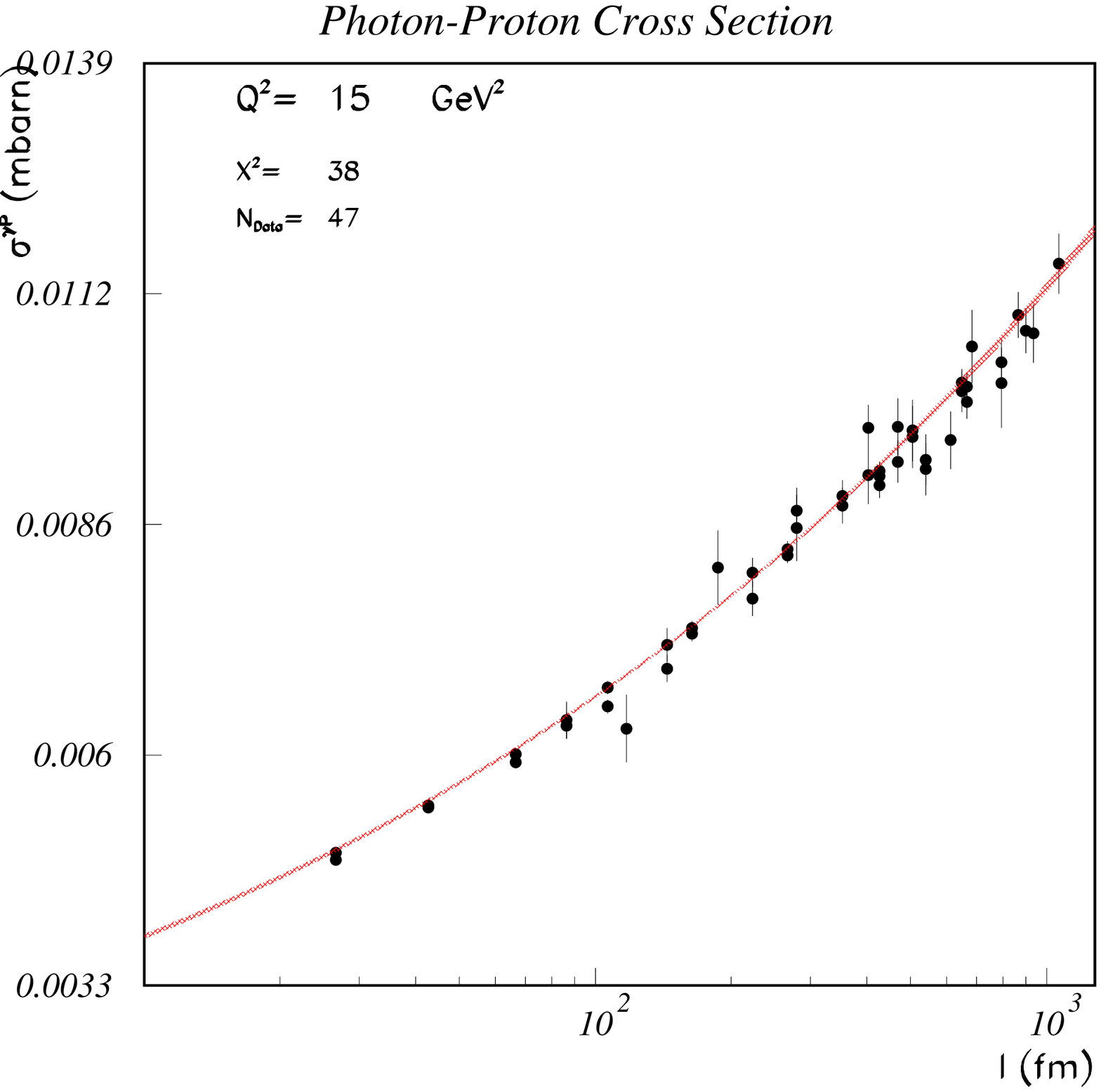}
   \includegraphics[width=0.45\textwidth]{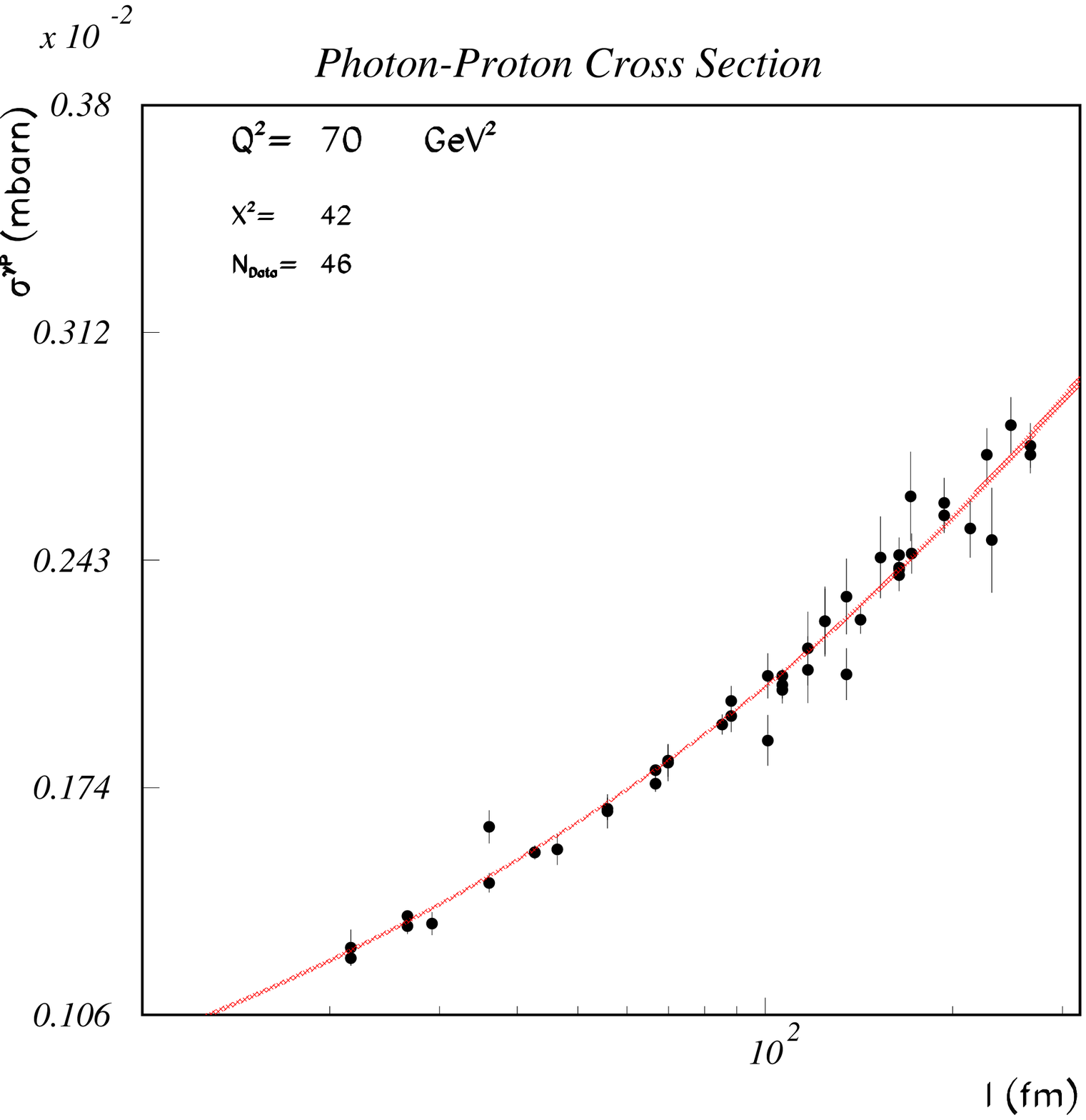}
\caption{\it Fits of the data using the function given in (\ref{eq:fitform}) for selected values of $Q^2$.  The values of $Q^2$ are given in the panels, as well as the number of data points fit and the value of $\chi^2$ using the best-fit parameters.  The band indicates the 68~\% credible interval from the fit function. The error bars on the points are from adding the statistical and systematic uncertainties in quadrature.}
\label{fig:results}
\end{center}
\end{figure}

The bin-centered data were then fit to the simple form
\begin{equation}
\label{eq:fitform}
\sigma(l,Q^2) = \sigma_1(Q^2) \left(\frac{l}{1{\rm \; fm}}\right)^{\lambda_{\rm eff}(Q^2)} 
\end{equation}
for fixed values of $Q^2$.  Some sample fits are shown in Fig~\ref{fig:results}.  These two-parameter fits were universally good, indicating that at fixed $Q^2$ values, the data are completely consistent with a simple power law behavior.  

\begin{figure}[hbpt]
\begin{center}
   \includegraphics[width=0.6\textwidth]{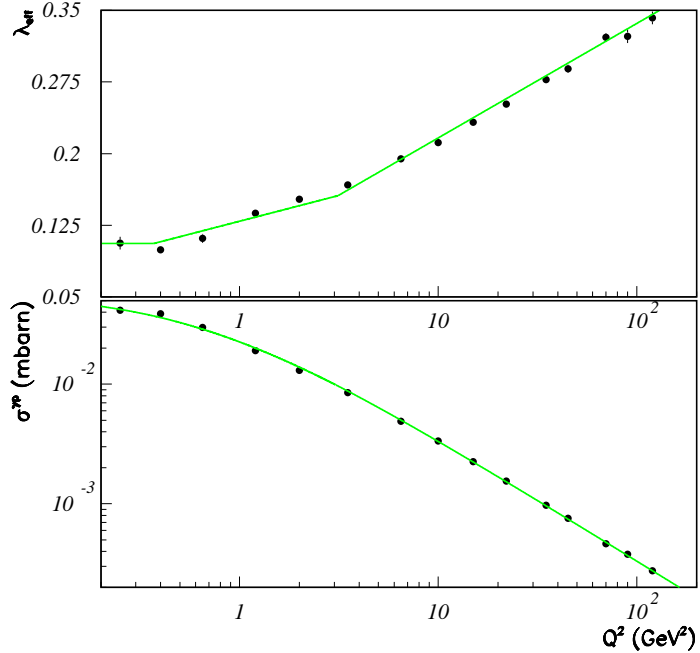}
\caption{\it Values of the parameters $\lambda_{\rm eff}$ and $\sigma^{\gamma p}(l=1 \; {\rm fm}) = \sigma_1$ (see Eq.~\ref{eq:fitform}) as a function of $Q^2$. The values expected from the global fit to the data (see Eq.~\ref{eq:global}) is shown as the green lines.  The uncertainties on the fit parameters are typically smaller than then size of the symbols.}
\label{fig:lambda}
\end{center}
\end{figure}

The parameters resulting from  these fits are shown in Fig.~\ref{fig:lambda}, together with the expectation from the formula used in the global fit of the data.  It is seen that the global fit does a decent job or representing the data, in particular the $Q^2$ dependence.  The dependence on the coherence length has the expected features: at small $Q^2$, the exponent $\lambda_{\rm eff} \approx 0.1$, a value compatible with the results from hadron-hadron scattering data.  For values above a few GeV$^2$, the exponent $\lambda_{\rm eff} \propto \ln(Q^2)$, with a transition region in between

\begin{figure}[hbpt]
\begin{center}
   \includegraphics[width=0.8\textwidth]{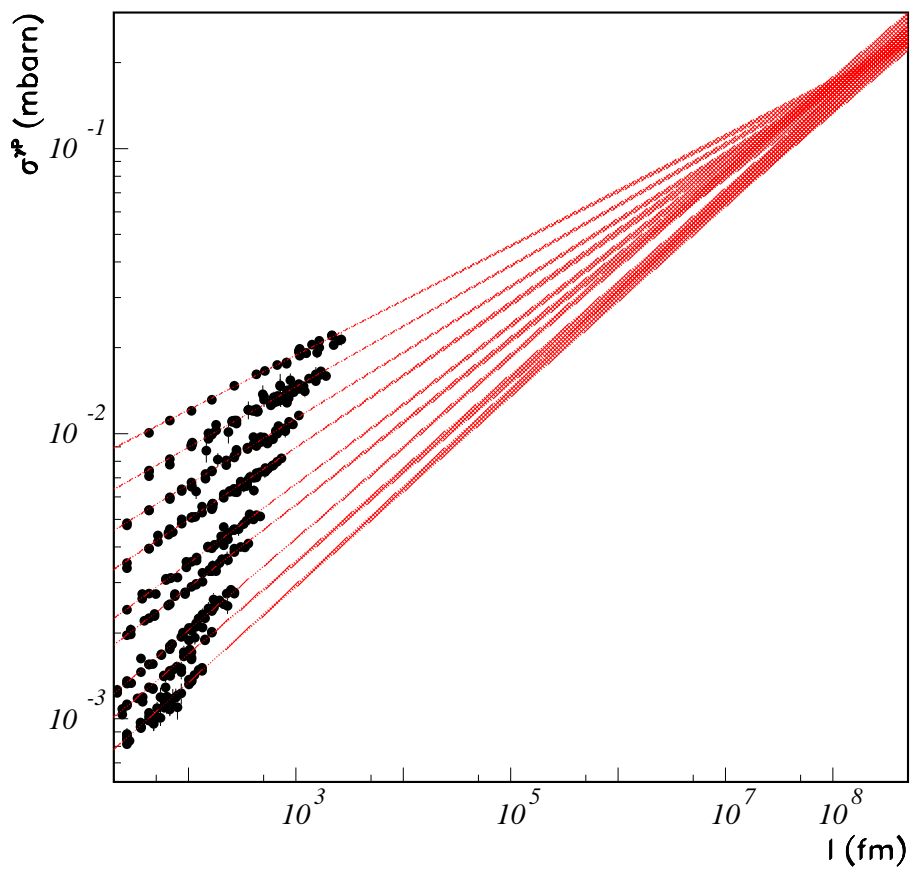}
\caption{\it Extrapolation of the fit functions to larger $l$ for values of $Q^2$ in the range $3.5\leq Q^2 \leq 90$~GeV$^2$. The width of the bands represent the 68~\% credible intervals for the functions from the data fits described in the text. The values of $Q^2$ range from $3.5$~GeV$^2$ (top curve) to $90$~GeV$^2$ (bottom curve) with the intermediate curves corresponding to intermediate $Q^2$ values.}
\label{fig:extrap}
\end{center}
\end{figure}

From Fig.~\ref{fig:lambda}, we note that while the cross section at $l=1$~fm decreases approximately as $1/Q^2$ for the larger values of $Q^2$, the dependence on the coherence length becomes steeper with increasing $Q^2$.  An extrapolation of the cross sections will result in the cross section for high $Q^2$ photons eventually becoming larger than the cross section for lower $Q^2$ photons.  We study this for $Q^2>3$~GeV$^2$, where the data in Fig.~\ref{fig:lambda} indicate that we have reached the simple scaling behavior of the cross section. An extrapolation of the cross sections is shown in Fig.~\ref{fig:extrap} and reveals that for $l\approx 10^8$~fm, the extrapolations merge.

\section{Discussion}

At small values of the coherence length, the photon-proton scattering cross section scales roughly as $1/Q^2$, as expected from color transparency.  However, as the coherence length increases, the scattering cross sections increase at different rates, and the scattering cross section for an initially small photon configuration will grow to be comparable to that for a smaller $Q^2$, or larger, photon configuration unless the growth of the cross section is modified.  If the photon states become comparable in size, they should evolve in the same way so that the cross sections evolve uniformly with coherence length, independently of their initial size, and with the same energy dependence as typical for hadronic cross sections.  This hadron-like energy dependence is observed for the smallest $Q^2$ values investigated (the energy dependence is the same as for hadron-hadron scattering).  We therefore expect that the slope of the cross section with coherence length should flatten as a function of coherence length; this could be an indication for saturation of the parton densities.  It is also possible that the power law behavior of the cross section changes with coherence length without saturation~\cite{ref:KLR} but this would again imply an interesting change in gluon dynamics.  The new effects should set in below $l=10^8$~fm to avoid the cross section crossing.  Given the relation given in Eq.~\ref{eq:coherence}, this means that a change should set in for $x>10^{-9}$.  First signs of the change in slope would presumably set in considerably earlier.  The approach to saturation and a fundamentally new state of matter is therefore  perhaps within reach of next generation lepton-hadron colliders~\cite{ref:EIC,ref:LHeC,ref:VHEeP}, an exciting prospect.

\section{Acknowledgments}
I would like to thank Matthew Wing and Henri Kowalski for many interesting discussions concerning this analysis.


\begin{thebibliography}{9}


\bibitem{ref:DL}
 A.~Donnachie, P.V.~Landshoff,
 Phys.\ Lett.\ B{\bf 296}, 227 (1992).

\bibitem{ref:sigmagp}
  J.~Breitweg {\it et al.}  [ZEUS Collaboration],
  Eur.\ Phys.\ J. \  C{\bf 7}, 609 (1999).
  
\bibitem{ref:caldwell}
A.~Caldwell, `Behavior of sigma(gamma p) at Large Coherence Lengths', arXiv:0802.0769v1.

\bibitem{ref:Gribov}
V.~N.~Gribov, `Space-time description of the hadron interaction at high energies', arXiv:0006158v1.

\bibitem{ref:Hand}
L.N.~Hand, Phys.\ Rev.  {\bf 129}, 1834 (1963).


\bibitem{ref:LES}
L.~Stodolsky, Phys.\ Lett.\ B{\bf 325}, 505 (1994).

\bibitem{ref:BK}
J.~Bartels and H.~Kowalski, `Diffraction at HERA and the Confinement Problem',   Eur.\ Phys.\ J. \  C{\bf 19}, 693 (2001).

\bibitem{ref:HERAcombined}
H.~Abramowicz {\it et al.} [H1 and ZEUS Collaborations],  Eur.\ Phys.\ J.\ C {\bf 75} 1 (2015).

\bibitem{ref:E665}
  M.~Adams {\it et al.}  [E665 Collaboration],
  Phys.\ Rev.\  D{\bf 54}, 3006 (1996).

\bibitem{ref:NMC}
  M.~Benvenuti {\it et al.}  [NMC Collaboration],
  Nucl.\ Phys.\  B{\bf 483}, 3 (1997).



\bibitem{ref:transparency}
B.~Bl\"attel, G.~Baym, L.L.~Frankfurt, M.~Strikman,
Phys.\ Lett.\ B{\bf 304}, 1 (1993).


\bibitem{ref:BAT}
A.~Caldwell, D.~Kollar, K.~Kr\"oninger, 
Comput. \ Phys. \ Commun. {\bf180},  2197 (2009).

\bibitem{ref:EIC}
 A.~Deshpande, R.~Milner, R.~Venugopalan and W.~Vogelsang,
  Ann.\ Rev.\ Nucl.\ Part.\ Sci.\  {\bf 55}, 165 (2005);	A. Accardi et al., `Electron Ion Collider: The Next QCD Frontier - Understanding the glue that binds us all', arXiv:1212.1701.
 
 \bibitem{ref:LHeC}
P.~Newman and A.~Stasto, Nature Phys.\ {\bf 9} (2013 448;\\
LHeC Study Group, J.L.~Abelleira Fernandez et al., J.\ Phys.\ {\bf G~39} (2012) 075001.
  
  \bibitem{ref:VHEeP}
A.~Caldwell, M.~Wing, `VHEeP: A very high energy electron-proton collider based on proton-driven plasma wakefield acceleration', arXiv:1509.00235v1.
 Talk presented at DIS2015, Dallas, USA, April 2015.  In proceedings, PoS (DIS2015) 240.

\bibitem{ref:KLR}
H.~Kowalski, L.~N.~Lipatov, and D.~A.~Ross, `The Behaviour of the Green Function for the BFKL Pomeron with Running Coupling', arXiv:1508.05744.

\end{thebibliography}
\end{document}